\documentstyle[12pt]{article}
\begin{document}

\begin{titlepage}
\setcounter{footnote}0
\begin{center}
\hfill AIV-97/I\\
\hfill       hep-th/9710120\\
\vspace{0.3in}
\bigskip
\bigskip
\bigskip
{\Large \bf Domain Wall in MQCD and}\\
\bigskip
{\Large \bf Supersymmetric Cycles in}\\
\bigskip
\smallskip
{\Large \bf Exceptional Holonomy Manifolds}
\bigskip
\bigskip
\bigskip
\bigskip
\bigskip
\bigskip
\\{\Large Anastasia Volovich}
\footnote{e-mail address: nastya@itp.ac.ru, nastya@main.mccme.rssi.ru}
\\
\bigskip
\begin{flushleft}
{\it Department of Physics,$~$ Moscow State
 University, Vorobjevi gori, 119899;}\hfill\\
{\it Landau Institute for Theoretical Physics,
Kosygina, 2, 117940;}\hfill\\
{\it   Mathematical College, Independent
  University of Moscow, P.O. Box 68,121108;}
\hfill
\\{\it  Moscow, Russia }
\end{flushleft}
\bigskip
\bigskip
\bigskip
\bigskip
\bigskip
\bigskip
\bigskip
\end{center}
\begin {abstract}

It was conjectured by Witten that a BPS-saturated domain wall
exists in the M-theory fivebrane version of QCD (MQCD)
and can be represented as a supersymmetric three-cycle in
the sense of Becker et al with an appropriate asymptotic
behavior. We derive the differential equation which
defines an associative cycle in $G_2$ holonomy seven-manifold
corresponding to the supersymmetric three-cycle
and show that it contains a sum of the Poisson brackets.
We study solutions of the differential equation with prescribed
asymptotic
behavior.  \end {abstract}

\end{titlepage}

\section{Introduction}

  Recently, many interesting results about supersymmetric
gauge theories in different dimensions have been obtained
by embedding gauge theories in string theory and
using properties of the latter to study the former.
D-branes has provided an arena to exchange results of
these theories and to advance both of them.
Three and four dimensional theories have been studied by using
configurations of branes in M-theory \cite{HW}.

   Witten used this method to explore the minimal  $N=1$
model with an $SU(n)$ vector multiplet in four dimensions \cite{Witten}.
He showed how for this model some of the outstanding properties of the
ordinary QCD such as confinement, a mass gap and spontaneous breaking of
a
discrete chiral symmetry can be approached from M-theory point of view.
A
configuration arising from $n$ Dirichlet fourbranes suspended between
two NS
fivebranes was used to study this theory.  It was shown that the
obtained
theory (MQCD) is confining, it has strings or flux tubes and undergoes
spontaneous breaking of the $Z_{n}$ chiral symmetry.

  A consequence of the spontaneously broken chiral symmetry
is that there can be a domain wall separating different vacua.
 Witten has suggested that a BPS-
saturated domain wall exists in MQCD and that it can be represented
as a supersymmetric three-cycle in the sense of Becker {\it et al}
\cite{Becker} with a prescribed
asymptotic behavior. BPS-saturated domain walls in four
dimensional supersymmetric field theories
have been considered in \cite{Shifman}.

The aim of this note is to study the Witten conjecture
on domain walls in MQCD.
First we will derive a differential equation for the
supersymmetric three-cycle. It turns out that this equation
contains
a sum of the Poisson brackets. Then we investigate solutions of
this equation.

\section{Domain Wall in MQCD}

  Domain walls separate spatial domains containing different
vacua. In \cite{Witten} it was suggested that domain walls exist in
supersymmetric gauge theory obtained by using a brane configuration
in M-theory.
 The domain wall is described \cite{Witten} as an M-theory
fivebrane of the form $ {\bf {R}}^{3} \times \cal S$ where $ {\bf
{R}}^{3} $
is parameterized by $x^{0}, x^{1}, x^{2} $ and $\cal S$ is
a three-surface in the seven manifold $ \tilde{Y} = {\bf {R}} \times Y$.
Here ${\bf R}$ is the copy of $x^{3}$ direction and
$Y = {\bf {R}}^{5} \times S^{1 }$. Near
$x^{3}=- \infty $,
$\cal S $ should look like ${\bf R} \times \Sigma$,
where $\Sigma $ is the Riemann surface defined by
$w = \zeta {\upsilon }^{-1}$, $t={\upsilon}^{n}$.
Here $w, \upsilon, t$ are complex coordinates
in ${\bf R}^{5} \times S^{1}$ with coordinates
$x^{4}$, $x^{5}$, $x^{7}$, $x^{8}$, $x^{6}$, $x^{10}$ :
$ \upsilon =x^{4}+ix^{5}$, $w=x^{7}+ix^{8},$
$s=x^{6}+ix^{10},$ $ t=e^{-s},$ $ 0 \leq x^{10} \leq 2\pi,$
$\zeta$ is a constant.
Near  $x^{3}=+\infty,~\cal S $ should look like
${\bf R} \times {\Sigma}^{'}$, where ${\Sigma}^{'}$
is the Riemann surface of an "adjacent" vacuum,
defined by
$w=exp({2 \pi i}/{n}) \zeta {\upsilon}^{-1}, t={\upsilon}^{n}. $
We have to find such $\cal S$ which is a supersymmetric three-cycle
with the described asymptotic behavior.
First we will derive a differential equation for the
supersymmetric three-cycle. It turns out that this equation
contains a sum of the Poisson brackets. Then we study  solutions
of this equation.

\section{Supersymmetric Cycle in $G_{2}$ Holonomy Manifolds}

A supersymmetric cycle is defined by the property that the
world-volume theory of a brane wrapping around it is supersymmetric.
The three-cycle is supersymmetric if the global supersymmetry
transformation can be undergone by $\kappa$-transformation, which
implies that
$$
P\_\Psi=\frac{1}{2}(1-\frac{i}{3!}\epsilon^{\alpha\beta\gamma}
\partial_{\alpha} X^{M} \partial_{\beta} X^{N} \partial_{\gamma} X^{P}
\Gamma_{MNP})\Psi=0,
$$
where $\Psi $ is covariantly constant ten-dimensional
spinor, $\Gamma_{MNP}$ are ten-dimensional $\Gamma$ matrices.

The conditions for the supersymmetric cycles in Calabi-Yau
3-folds have been analyzed on \cite{Becker1}. It was shown that
a supersymmetric three-cycle is one for which the pullback of
K\"{a}hler form  $J$ vanishes and the pullback of the holomorphic
3-form $\Omega$ is a constant multiple of volume element, namely
$\ast X(J) = 0,~~\ast X(\Omega) \sim 1$,
where $X(.)$ denotes the pullback and $\ast$ is a Hodge dual
on membrane world-volume.
In \cite{Becker} it was mentioned that
supersymmetric cycles may be complex or special Lagrangian submanifolds
and
also of exceptional type which exist in $Spin(7)$, $SU(4)$  and $G_{2}$
holonomy manifolds. To study supersymmetric cycles one uses the concept
of
calibration \cite{Becker}. A calibration is a closed $p$-form $\Phi$ on
a
Riemannian manifold of dimension $n$ such that its restriction to each
tangent $p$-plane of $M$ is less or equal to the volume of the plane.
Submanifolds for which there is equality are said to be calibrated by
$\Phi$.

In the case of domain walls in MQCD \cite{Witten} one deals with
a seven dimensional flat manifold of $G_{2}$ holonomy and with the
associative calibration. $G_{2}$ is a subgroup of $O(7)$ which leaves
the
constant spinor invariant. Just as in Calabi-Yau case in the case of
$G_{2}$
holonomy manifold there is a canonical 3 form $\Phi$ and its Hodge dual
$\ast \Phi$ which are covariantly constant \cite{Joyce}.  If we choose
the local veilbein
so that the metric is $\sum_{i=1}^{n} e_{i} \otimes e_{i}$ ,
locally the $G_{2}$ invariant 3-form $\Phi$ can be written as
\cite{Vafa}
with
\begin{equation}
\Phi=e_{1} \wedge e_{2} \wedge e_{7}
+ e_{1} \wedge e_{3} \wedge e_{6}
+ e_{1} \wedge e_{4} \wedge e_{5}
+ e_{2} \wedge e_{3} \wedge e_{5}
- e_{2} \wedge e_{4} \wedge e_{6}
+ e_{3} \wedge e_{4} \wedge e_{7}
+ e_{5} \wedge e_{6} \wedge e_{7}.
\label{form}
\end{equation}

 A supersymmetric three-cycle $\cal S$ in $\tilde{Y}$ is one for which
the
pullback $\Phi | _{\cal S}$ of this three-form is a constant multiple of
the volume element.
This condition is written below as a partial differential equation
in terms of Poisson brackets.

\section{Differential equation for a supersymmetric cycle}

Let us derive a differential equation defining a supersymmetric
cycle ${\cal S}$.

\bigskip

{\it  Introducing notations.}

\bigskip

Let us choose  $\upsilon, \bar {\upsilon}, x^{3}$ as coordinates on $
\cal S$. The 3-cycle
$\cal S $ is embedded in ${\bf R}^{7}$ as

\begin{equation}
w=w(x^{3}, \upsilon, \bar{\upsilon}), ~~~~~
s=s(x^{3}, \upsilon, \bar{\upsilon}),
\label{2}
\end{equation}

$$\upsilon =x^{4}+ix^{5},~~~w=x^{7}+ix^{8},~~~s=x^{6}+ix^{10}.$$

Here functions $s$ and $w$ have asymptotic behavior of the form

\begin{equation}
w(+\infty,\upsilon,\bar{\upsilon})=
e^{\frac{2\pi i}{n}}\zeta {\upsilon}^{-1},~~~
s(+\infty,\upsilon,\bar{\upsilon})=-n \ln{\upsilon},
\label{3}
\end{equation}
and
\begin{equation}
w(-\infty,\upsilon,\bar{\upsilon})=
\zeta {\upsilon}^{-1}, ~~~
s(-\infty,\upsilon,\bar{\upsilon})=-n \ln{\upsilon},
\label{4}
\end{equation}

We denote $x^{6}=A,~x^{7}=C,~x^{8}=D,~x^{10}=B$
with $A,B,C$ and $D$ being functions of $x^{3}, x^{4}, x^{5}$.

Let us set
\begin{equation}
e_{1}=dx^{10},
e_{2}=dx^{5},
e_{3}=dx^{3},
e_{4}=dx^{7},
e_{5}=dx^{4},
e_{6}=dx^{6},
e_{7}=dx^{8},
\label{ne}
\end{equation}
and denoted $x^{4}=y_{1},~x^{5}=y_{2},~x^{3}=y_{3}$
then the form (\ref{form}) will take the form

$$\Phi= dB \wedge dy_{2} \wedge dD+
   dB \wedge dy_{3} \wedge dA +dB \wedge dC \wedge dy_{1} -
   dy_{2} \wedge dC \wedge dA$$

  \begin{equation}
+dy_{3} \wedge dC \wedge dD +dy_{1} \wedge dA \wedge dD+
dy_{2} \wedge dy_3 \wedge dy_1.
\label{Phi}
\end{equation}

\bigskip

{\it  Differential equation and
boundary conditions.}
\bigskip

{\bf Lemma A.} The surface ${\cal S}$ embedded in ${\bf R}^{7}$ as
(\ref{2}) is a
supersymmetric three-cycle if
$A, B, C, D$ as functions of $x^{3}, x^{4}, x^{5}$
satisfy the following non-linear partial
differential equation
$$
1+\{A,B\}_{(1,2)}
+\{C,D\}_{(1,2)}
+\{B,D\}_{(1,3)}+$$
\begin{equation}
\{B,C\}_{(2,3)}
+\{C,A\}_{(1,3)}
+\{A,D\}_{(2,3)}=\sqrt{g},
\label{EE}
\end{equation}
where we denoted $x^{4}=y_{1},~x^{5}=y_{2},~x^{3}=y_{3}$ and
defined the Poisson brackets as
$$
\{f, g\}_{(i,j)}=
\frac{\partial f}{\partial y_{i}}
\frac{\partial g}{\partial y_{j}}-
\frac{\partial g}{\partial y_{i}}
\frac{\partial f}{\partial y_{j}}.
$$
Here $f=f(y_{1},y_{2},y_{3})~$,
$g=g(y_{1},y_{2},y_{3})$,  $~i,j= 1,2,3$ and $g$
is the determinant of the induced metric
\footnote{In the previous version of this paper  the determinant
was set equal to 1. This lead to an extra condition for the solution.
I am grateful to  J. Maldacena for pointing out this fact.}

\begin{equation}
g_{ij}=\delta_{ij}+\partial _{i}A\partial _{j}A+
\partial _{i}B\partial _{j}B+
\partial _{i}C\partial _{j}C+
\partial _{i}D\partial _{j}D
\label{mat}
\end{equation}

We will consider here the case $n=2$ and $\zeta$ being a real number.
Then the boundary conditions (\ref{3}),(\ref{4}) will take the form:

$$C(y_{1},y_{2},\pm \infty)=\pm c(y_{1},y_{2}), ~~~~~
c(y_{1},y_{2})=
- \frac{y_{1}\zeta}{{y_{1}}^{2}+{y_{2}}^{2}}.$$

$$ D(y_{1},y_{2},\pm \infty)=
\pm d(y_{1},y_{2}),~~~~~
d(y_{1},y_{2})=
\frac{y_{2}\zeta}{{y_{1}}^{2}+{y_{2}}^{2}},~~ $$

$$A(y_{1},y_{2},\pm \infty)=
a(y_{1},y_{2}),~~~~~
a(y_{1},y_{2})=-\ln{({y_{1}}^{2}+{y_{2}}^{2})},$$

\begin{equation}
B(y_{1},y_{2},\pm \infty)=b(y_{1},y_{2}),~~~~~
b(y_{1},y_{2})=-2
\arctan{\frac{y_{2}}{y_{1}}}+\pi.
\label{10}
\end{equation}

{\bf Proof.}
The equation defining a supersymmetric three-cycle ${\cal S}$
is \cite{Becker}
\begin{equation}
\Phi | _{{\cal S}} = \sqrt{|g|}dx^{3} \wedge dx^{4} \wedge
dx^{5}.
\label{vac}
\end{equation}
 Since we set (\ref{ne})
the form (\ref{form})  takes the form of a sum of the Poisson brackets

$$\Phi= (
1+\{A,B\}_{(1,2)}
+\{C,D\}_{(1,2)}
+\{B,D\}_{(1,3)}+$$
\begin{equation}
\{B,C\}_{(2,3)}
+\{C,A\}_{(1,3)}
+\{A,D\}_{(2,3)})
dy_{1} \wedge dy_2 \wedge dy_3.
\label{phi-pb}
\end{equation}
The condition for the surface to be a supersymmetric three-cycle
is reduced to (\ref{EE}).

\bigskip

{\it  Vacuum solutions.}
\bigskip

Let us check that the vacuum configurations satisfy equation (\ref{EE}).
\bigskip

{\bf Lemma B.} The surface ${\cal S}_{0}$ embedded in ${\bf R}^{7}$ as
(\ref{2}) with no dependence on $x_{3}$ and analytical functions $w$
and $s$, i.e. $w=w(v)$, $s=s(v)$ is a
supersymmetric three-cycle.
{\bf Proof.}
We denote $A$, $B$ and  $C$, $D$ defining ${\cal S}_{o}$
as $A=a,$ $B=b$, $C=c$ and $D=d$. They satisfy the
Cauchy-Riemann conditions
\begin{equation}
\label{CR}
\partial _1 a =\partial _2 b,~~~~~\partial _2a=-\partial _1b.
\end{equation}
For ${\cal S}_{0}$ we have

\begin{equation}
\Phi|_{{\cal S}_{0}}=
(1+\{a,b\}_{(1,2)}+\{c,d\}_{(1,2)})dy_{1} \wedge dy_2 \wedge dy_3.
\label{phi}
\end{equation}
Due to the Cauchy-Riemann conditions (\ref{CR})
we have
\begin{equation}
1+\{a,b\}_{(1,2)}+\{c,d\}_{(1,2)}=
1+(\partial _{1}a)^{2}+(\partial _{2}a)^{2}+
(\partial _{1}c)^{2}+(\partial_{2}c)^{2}.
              \label{PB0}
\end{equation}
The induced metric on
${\cal S}_{0}$ is
\begin{equation}
(g_{ik})|_{{\cal S}_{0}}=
\left(
 \begin{array}{ccc}
  g^{(0)}_{11}
   ~~&~~0~~ & 0 \\
 0~~&~~
g^{(0)}_{11}~~&~~0 \\
       0 &0& 1
\end{array}
\right),~~~~~g^{(0)}_{11}=1+(\partial _{1}a)^{2}+(\partial _{2}a)^{2}+
   (\partial _{1}c)^{2}+(\partial_{2}c)^{2}.
\end{equation}
The square root of the determinant of the induced metric on
${\cal S}_{0}$ is
$$\sqrt{g}|_{{\cal S}_{0}}=
1+(\partial _{1}a)^{2}+(\partial _{2}a)^{2}+
     (\partial _{1}c)^{2}+(\partial_{2}c)^{2},
$$
and it coincides with (\ref{PB0}) , that proves the Lemma B.

\bigskip

{\it  Ansatz.}
\bigskip

To solve equation (\ref{EE}) with boundary conditions (\ref{10})
we will use an expansion into series over the powers of the function
$${\gamma}(y_{3})=\frac{1}{e^{y_{3}}+e^{-y_{3}}},$$
which vanishes in the limit $y_{3}=x^{3}\to \pm \infty$.
This function satisfies $|\gamma (x)|<1/2$.

The following ansatz will be used

\begin{equation}
A=a(y_{1},y_{2})+\sum_{k=1}^{\infty} {\gamma}^{2k}(y_{3})
a_{2k}(y_{1},y_{2}),
\label{11a}
\end{equation}

\begin{equation}
B=b(y_{1},y_{2})+\sum_{k=1}^{\infty} {\gamma}^{2k}
(y_{3})b_{2k}(y_{1},y_{2}),
\label{11b}
\end{equation}

\begin{equation}
C=\beta(y_{3})c(y_{1},y_{2}),
\label{11c}
\end{equation}

\begin{equation}
D=\beta(y_{3})d(y_{1},y_{2}),
\label{11d}
\end{equation}
here
$$ \beta(y_{3})=\frac{e^{y_{3}}-e^{-y_{3}}}{e^{y_{3}}+e^{-y_{3}}}.$$

The boundary conditions (\ref{10}) are
satisfied for the above ansatz. In fact, the ansatz
(\ref{11a})-(\ref{11d})
can be used for arbitrary functions $a, b, c, d$ satisfying
the Cauchy-Riemann conditions.
In order to solve equations (\ref{EE}) we have to find functions
$a_{2k}(y_{1},y_{2})$, $b_{2k}(y_{1},y_{2})$.
Let us mention that in principle one can add to (\ref{11c})
and (\ref{11d}) series over $\gamma (y_{3})$
as we did in (\ref{11a}) and (\ref{11b}). We fix here this arbitrariness
to simplify calculations.

\bigskip

{\it Calculations.}
\bigskip

Due to a simple differential algebra

\begin{equation}
{\partial}_{3} \beta = 4 {\gamma}^{2}, ~~~
  {\partial}_{3} {\gamma}^{k}= -k\beta{\gamma}^{k},~~~
  {\beta}^{2}=1-4{\gamma}^{2},
\label{da}
\end{equation}
we have the following expressions for combinations of  Poisson brackets
for
the ansatz (\ref{11a})-(\ref{11d}):

\begin{equation}
\{C,A\}_{(1,3)}+ \{A,D\}_{(2,3)} =4\gamma ^{2}(-c\partial _{1}A
+d\partial _{2}A),
\label{PB1}
\end{equation}

\begin{equation}
\{B,D\}_{(1,3)}+ \{B,C\}_{(2,3)} =4\gamma ^{2}(c\partial _{2}B
+d\partial _{1}B).
\label{PB2}
\end{equation}

Therefore, representing
$$
A=a+{\cal A},~~~~~~~B=b+{\cal B}$$
one can write
$$
\{C,A\}_{(1,3)}+ \{A,D\}_{(2,3)}+ \{B,D\}_{(1,3)}+ \{B,C\}_{(2,3)} =
$$
\begin{equation}
\label{PBS}
4\gamma ^{2}(-c(\partial _{1}{\cal A} -\partial _{2}{\cal B})
+d(\partial _{2}{\cal A}+\partial _{1}{\cal B})),
\end{equation}
i.e. this combination of the Poisson brackets starts from the 4-th order
on $\gamma$,
$$
\{C,A\}_{(1,3)}+ \{A,D\}_{(2,3)}+ \{B,D\}_{(1,3)}+ \{B,C\}_{(2,3)} =
$$
$$
4\gamma ^{2}\sum _{k=1}^{\infty}\gamma ^{2k}[-c(\partial _{1}a_{2k}
-\partial _{2}b_{2k} )
+d(\partial _{2}a_{2k}+\partial _{1}b_{2k})]
$$
One also has

$$\{A,B\}_{(1,2)}+\{C,D\}_{(1,2)}=
(\partial _{1}a)^{2}+(\partial _{2}a)^{2}
+\beta ^{2} ((\partial _{1}c)^{2}+(\partial _{2}c)^{2})+
$$

\begin{equation}
\label{PB5}
\partial _{1}a\partial _{2} {\cal B}
+\partial _2 b\partial _{1} {\cal A}
-\partial _{2}a\partial _{1} {\cal B}
-\partial _1 b\partial _{2} {\cal A}+\{ {\cal A},{\cal B}\}
\end{equation}
We see that only the powers of $\gamma^2$ enter the expression
$$
{\cal P}\equiv 1+\{C,D\}_{(1,2)}+\{A,B\}_{(1,2)}
+
\{C,A\}_{(1,3)}+ \{A,D\}_{(2,3)}+ \{B,D\}_{(1,3)}+ \{B,C\}_{(2,3)}
$$
and  in the lowerest orders on $\gamma^2$  we have

$$
{\cal P}=
g_{11}^{(0)}-4\gamma ^{2}((\partial _{1}c)^{2}+(\partial_{2}c)^{2})+
\gamma ^{2}[\partial _{1}a(\partial _{2}b_{2}+\partial _{1}a_{2})
-\partial _{2}a(\partial _{1}b_{2}-\partial _{2}a_{2})]+
$$
\begin{equation}
\label{lo}
+\gamma ^{4}[\partial _{1}a(\partial _{2}b_{4}+\partial _{1}a_{4})
-\partial _{2}a(\partial _{1}b_{4}-\partial _{2}a_{4})+
4c(-\partial _{1}a_{2}+\partial _{2}b_{2})
+4d(\partial _{2}a_{2}+\partial _{1}b_{2})
\end{equation}
$$
+\partial _{1}a_{2}\partial _{2}b_{2}-
\partial _{2}a_{2}\partial _{1}b_{2}]
$$

Let us now examine the form of the determinant. Under ansatz
(\ref{11a})-(\ref{11d}) we have

$$
\left(
 \begin{array}{ccc}
   1+\Delta_{11}^{(0)}+\gamma ^{2}\Delta^{(1)}_{11}
   +\gamma ^{4}\Delta^{(2)}_{11}+...&
   \gamma ^{2}\Delta^{(1)}_{12}+ \gamma^{4}\Delta^{(2)}_{12}
       & \beta \gamma ^{2}(
       \Delta_{13}^{(1)}+\gamma ^{2}\Delta_{13}^{(2)}+...) \\
 \gamma ^{2}\Delta^{(1)}_{12}+ \gamma^{4}\Delta^{(2)}_{12}+...&
 1+\Delta_{11}^{(0)}+\gamma ^{2}\Delta^{(1)}_{22}
   +\gamma ^{4}\Delta^{(2)}_{22}+...&
 \beta\gamma  ^{2}(\Delta_{13}^{(1)}+\gamma ^{2}\Delta_{13}^{(2)}+...)
\\
 \beta\gamma  ^{2}(\Delta_{23}^{(1)}+\gamma ^{2}\Delta_{23}^{(2)}+...)
 &\beta\gamma  ^{2}(\Delta_{13}^{(1)}+\gamma ^{2}\Delta_{13}^{(2)}+...)
 & 1+\gamma^{4}\Delta_{33}^{(2)}+...
 \end{array}
 \right)
 $$
\begin{equation}
\Delta_{11}^{(1)}=2\partial _{1}a\partial _{1}a_{2} +2\partial
_{1}b\partial _{1}b_{2} -4(\partial _{1}c)^{2} -4(\partial _{2}c)^{2},
 \end{equation}

\begin{equation}
\Delta_{22}^{(1)}=2\partial _{2}a\partial _{2}a_{2}
+2\partial _{2}b\partial _{2}b_{2}
-4(\partial _{1}c)^{2}
-4(\partial _{2}c)^{2},
\end{equation}

\begin{equation}
\Delta_{12}^{(1)}=
\partial _{1}a_{2}\partial _{2}a +
\partial_{1}a\partial _{2}a_{2} +
\partial _{1}b_{2}\partial _{1}a
-\partial_{2}a\partial _{2}b_{2},
 \end{equation}

$$
\Delta_{13}^{(1)}=
4(c\partial _{1}c+d\partial _{1}d)-2(a_{2}\partial _{1}a-
b_{2}\partial _{2}a),
$$

\begin{equation}
\Delta_{23}^{(1)}=
4(c\partial _{2}c+d\partial _{2}d)-2(a_{2}\partial _{2}a+
b_{2}\partial _{1}a),
 \end{equation}
\begin{equation}
\Delta_{33}^{(2)}=
 16(c^{2}+d^{2})+4(a_{2}^{2}+b_{2}^{2}),
\end{equation}

\begin{equation}
\Delta_{11}^{(2)}=
2\partial _{1}a\partial _{1}a_{4} +
(\partial_{1}a_{2})^{2} +
2\partial _{1}b_{1}\partial _{1}b_{4} +
(\partial_{1}b_{2})^{2},
 \end{equation}

\begin{equation}
\Delta_{22}^{(2)}=
2\partial _{2}a\partial _{2}a_{4} +
(\partial_{2}a_{2})^{2} +
2\partial _{2}b_{1}\partial _{2}b_{4} +
(\partial_{2}b_{2})^{2}.
 \end{equation}

Performing calculation of the determinant
and expanding the result
over powers of $\gamma$ we see that it contains only the powers of
$\gamma^{2}$ and  does not contain the terms proportional to $\beta$.
We have

\begin{equation}
g=(g_{11}^{(0)})^{2}+\gamma ^{2}g_{11}^{(0)}\rho _{2}+
\gamma ^{4}\rho _{4}+....
\end{equation}
where
$$\rho _{2}=\Delta_{11}^{(1)}
+\Delta_{22}^{(1)},$$

$$\rho_{4}=\Delta_{11}^{(1)}\Delta_{22}^{(1)}-
\Delta_{12}^{(1)}\Delta_{12}^{(1)}
 -g^{(0)}_{11}(\Delta_{13}^{(1)}\Delta_{13}^{(1)}+
\Delta_{23}^{(1)}\Delta_{23}^{(1)})
+g_{11}^{(0)}(\Delta_{11}^{(2)}+\Delta_{22}^{(2)})
+\Delta_{33}^{(2)}(g_{11}^{(0)})^{2}.$$

The condition for the surface to be a supersymmetric three-cycle will
lead
us to
the following equation
\begin{equation}
{\cal P}^{2}=g.
\label{Pg}
\end{equation}

Let us require vanishing of the coefficients in front of $\gamma^{k}$
for all $k$. This will lead us to the system of  recursive
differential equations
for $a_{2k}$, $b_{2k}.$
The coefficients in front of ${\gamma}^2$ in the right and left
hand sides of
(\ref{Pg}) vanish identically.
In the fourth order we get an equation in which the terms
with $a_{4}$ and $b_{4}$ are canceled and
in the fourth order we left with nonlinear equation for
$a_{2}$  and $b_{2}$. We can take one of them equal to zero or assume
some
relation between them.

Let us  assume the Cauchy-Riemann
conditions on $A$ and $B$ as  functions  of $y_{1},$ $y_{2}$.
>From  equation
(\ref{PBS}) we see that if one assumes the Cauchy-Riemann
conditions for $A$ and $B$ we will get that

\begin{equation}
\label{PB-CR}
{\cal P}|_{CR}=1+(\partial _{1}A)^{2}
+(\partial _{2}A)^{2}
+\beta^{2}((\partial _{1}c)^{2}+(\partial _{2}c)^{2}).
\end{equation}

The determinant also simplifies.
We have
\begin{equation}
\label{g11}
g_{11}=g_{22}=1+(\partial _{1}A)^{2}+(\partial _{2}A)^{2}+
(\partial _{1}C)^{2}+(\partial _{2}C)^{2},
\end{equation}
\begin{equation}
\label{g12}
g_{12}=g_{21}=\partial _{1}A\partial _{2}A+
\partial _{1}B\partial _{2}B +
\partial _{1}C\partial _{2}C+
\partial _{1}D\partial _{2}D=0,
\end{equation}

$$g_{13}=g_{31}=\partial _{1}A\partial _{3}A+
\partial _{1}B\partial _{3}B +
\partial _{1}C\partial _{3}C+
\partial _{1}D\partial _{3}D,$$

$$g_{23}=g_{32}=\partial _{2}A\partial _{3}A+
\partial _{2}B\partial _{3}B +
\partial _{2}C\partial _{3}C+
\partial _{2}D\partial _{3}D.$$

We notice that
\begin{equation}
\label{P-g11}
{\cal P}=g_{11}=g_{22},
\end{equation}
and the metric has the form
\begin{equation}
(g_{ik})|_{CR}=
\left(
 \begin{array}{ccc}
   g_{11}~~&~~0~~ &~~g_{13} \\
        0~~&~~ g_{11}~~&~~g_{23} \\
        g_{31}~~ &~~g_{32}~~&~~ g_{33}
 \end{array}
\right).
\label{g-PK}
\end{equation}
Therefore
\begin{equation}
det(g_{ik})|_{CR}=
   g_{11}(g_{11}g_{33}-(g^{2}_{13}+g^{2}_{23}))
\label{det-PK}
\end{equation}
Hence the equation (\ref{EE}) under the assumption of analyticity
has the form
\begin{equation}
g_{11}(g_{33}-1)=g^{2}_{13}+g^{2}_{23}.
\label{det-PK}
\end{equation}
or explicitly,
$$
[\partial _{1}A\partial _{3}A-
\partial _{2}A\partial _{3}B +
\partial _{1}C\partial _{3}C-
\partial _{2}C\partial _{3}D]^{2}+
[\partial _{2}A\partial _{3}A+
\partial _{1}A\partial _{3}B +
\partial _{2}C\partial _{3}C+
\partial _{1}C\partial _{3}D]^{2}=
$$
\begin{equation}
[1+(\partial _{1}A)^{2}+
(\partial _{2}A)^{2}+
(\partial _{1}C)^{2}+
(\partial _{2}C)^{2}]\cdot
[(\partial _{3}A)^{2} +
(\partial _{3}B)^{2}+
(\partial _{3}C)^{2}+
(\partial _{3}D)^{2}]
\label{exp}
\end{equation}

The form of equation (\ref{EE}) depends on the embedding
of the surface ${\cal S}$
and also on the choice of the basis (\ref{ne}).
We will consider now another nonlinear equation which is similar to
(\ref{EE}) and for which we prove the existence of a solution with the
prescribed asymptotic behavior as a series over  $\gamma$.

\section {Solution of a nonlinear equation}

Now instead of (\ref{ne}) we set
\begin{equation}
e_{1}=dx^{6},
e_{2}=dx^{4},
e_{3}=dx^{5},
e_{4}=dx^{7},
e_{5}=dx^{3},
e_{6}=dx^{8},
e_{7}=dx^{10},
\label{e}
\end{equation}
then the form (\ref{form}) will take the form

$$\Phi= dA \wedge dx^{4} \wedge dB+
   dA \wedge dx^{5} \wedge dD +dA \wedge dC \wedge dx^{3} +
   dx^{4} \wedge dx^{5} \wedge dx^{3}$$

\begin{equation}
-dx^{4} \wedge dC \wedge dD +dx^{5} \wedge dC \wedge dB+
dx^{3} \wedge dD \wedge dB.
\label{phin}
\end{equation}
We consider now the condition
$\Phi | _{\cal S} = dx^{3} \wedge dx^{4} \wedge
dx^{5}$. It can be written as
$$
\{A,B\}_{(3,2)}
+\{A,D\}_{(1,3)}
+\{A,C\}_{(1,2)}+$$
\begin{equation}
\{C,D\}_{(3,2)}
+\{C,B\}_{(3,1)}
+\{D,B\}_{(1,2)}=0,
\label{EEn}
\end{equation}

To solve equation (\ref{EEn}) with boundary conditions
(\ref{10}) we will use the same ansatz as below, see (\ref{11a})-
(\ref{11d}). We have
\begin{equation}
\{A,D\}_{(3,1)}=
-4d{\partial}_{1}a \cdot{\gamma}^{2}-
4{\gamma}^{2}d \cdot {\sum_{k=1}^{\infty}}
{\gamma}^{2k}{\partial}_{1} a_{2k}-
2(1-4{\gamma}^{2}) {\partial}_{1} d \cdot {\sum_{k=1}^{\infty}} k
{\gamma}^{2k}a_{2k},
\label{PB2}
\end{equation}
$$\{A,B\}_{(3,2)}=
-2 \beta (\sum_{k=1}^{\infty}k{\gamma}^{2k} a_{2k})
({\partial}_{2} b + \sum_{m=1}^{\infty}
{\gamma}^{2m}{\partial}_{2} b_{2m})  +$$
\begin{equation}
2 \beta (\sum_{k=1}^{\infty}k{\gamma}^{2k} b_{2k})
({\partial}_{2} a + \sum_{m=1}^{\infty}
{\gamma}^{2m}{\partial}_{2} a_{2m}),
\label{PB5}
\end{equation}
\begin{equation}
\{A,C\}_{(1,2)}=
\beta ({\partial}_{1} a {\partial}_{2} c -
{\partial}_{2} a {\partial}_{1} c)+
\beta \cdot \sum_{k=1}^{\infty}
{\gamma}^{2k}
({\partial}_{2} c
{\partial}_{1} a_{2k}-
{\partial}_{1} c
{\partial}_{2} a_{2k}),
\label{PB3}
\end{equation}

\begin{equation}
\{C,D\}_{(3,2)}=
4 \beta {\gamma}^{2}(c {\partial}_{2} d - d{\partial}_{2} c),
\label{PB6}
\end{equation}
where ${\partial}_{i}$ denotes $\frac{\partial}{\partial y_{i}}$,
$i=1,2.$ One gets $\{B,C\}_{(3,1)}$ from (\ref{PB2}) after the
substitution
$A \rightarrow B,~~a_{k} \rightarrow b_{k},$
$D \rightarrow C$
and $\{B,D\}_{(1,2)}$ from ({\ref{PB3}}) after
the substitution
$A \rightarrow B,~~a_{k} \rightarrow b_{k},$
$C \rightarrow D.$

Only terms $\beta {\gamma}^{2k}$ and
${\gamma}^{2k}$ contain the dependence on $y_{3}$.
Let us require vanishing of the coefficients
in front of $\beta {\gamma}^{2k}$ and
${\gamma}^{2k}$ for all $k$. This will lead us to the following
recursive system of linear equations for $a_{2k}(y_{1},y_{2})$,
$b_{2k}(y_{1},y_{2})$

\begin{equation}
-\partial _{2}c  \cdot a_{2k}+
 \partial _{1}c  \cdot b_{2k}=\frac{1}{2k}\cdot P_{k};
\label{le1}
\end{equation}

\begin{equation}
-\partial_{1} a \cdot a_{2k}
+\partial_{2} a \cdot b_{2k}-\frac{1}{2k}({\cal D}_1 a_{2k}
+{\cal D}_2 b_{2k}) =\frac{1}{2k}\cdot R_{k},
\label{le2}
\end{equation}
where ${\cal D}_i,$ $i=1,2$ are given by
\begin{equation} {\cal
D}_1=\partial_{1} c \cdot \partial_{2}- \partial_{2} c \cdot
\partial_{1},~~
{\cal D}_2=\partial_{1} c \cdot \partial_{1}+ \partial_{2} c \cdot
\partial_{2},
\label{D12}
\end{equation}
$P_{k}$ and $R_{k}$ depend only
on $a_{i},~b_{i},~c_{i},~d_{i}$ with $i \leq 2k-2,$ namely
\begin{equation}
P_{k}=
\delta_{k,1}P_1-
4(c{\partial}_{1}b_{2k-2}+
d{\partial}_{1}a_{2k-2})+
8(k-1)({\partial}_{1} c \cdot b_{2k-2}-
{\partial}_{2} c \cdot a_{2k-2}),
\label{Pk}
\end{equation}
$P_1=-4(c\partial_{1}b+
d\partial_{1}a)$ and
\begin{equation}
R_{k}=\delta _{k,1}R_1+2\cdot \sum_{s=1}^{k-1}s \cdot [a_{2s}
{\partial}_{2}b_{2k-2s}-
b_{2s}{\partial}_{2}a_{2k-2s}] ,
\label{Rk}
\end{equation}
$R_1=-4(c{\partial}_{2} d-d\partial _{2} c).$

Note that in our notations $a_{0}=b_{0}=0.$
If one finds $a_{2k}$ from (\ref{le1}) and
substitutes to (\ref{le2}) then due to the
Cauchy-Riemann conditions for $(a, b)$ and $(c, d)$ one gets
the equation containing only derivative over $y_{2}$:

\begin{equation}
({\partial}_{2}+V_{2k}) b_{2k} =L_{2k},
\label{ncl}
\end{equation}
where
\begin{equation}
V_{2k}= kW_1+W_0;
\label{v}
\end{equation}
$W_1$ and $W_0$ do not depend on $k$:
\begin{equation}
\label{w1}
W_{1}=\frac{ 2 ({\partial}_{1} a \partial_{1} c
-\partial_{2} a \partial_{2} c)}
{({\partial}_{1} c)^{2}+({\partial}_{2} c)^{2}}
;
\end{equation}
\begin{equation}
W_{0}=\frac{
{\partial}_{2}({\partial}_{1} c / {\partial}_{2} c)  \cdot
{\partial}_{1} c -
{\partial}_{1}({\partial}_{1} c / {\partial}_{2} c ) \cdot
{\partial}_{2} c }{({\partial}_{1} c)^{2}+({\partial}_{2} c)^{2}}
\cdot {\partial}_{2} c,
\label{w0}
\end{equation}
\begin{equation}
L_{2k}=
 \frac {\partial_{2} c}{(\partial_{1} c)^2+(\partial_{2}c)^2}
 \cdot
[-R_{k} + P_{k}\frac{\partial_{1} a}{\partial_{2} c}+
\frac{1}{2k} \frac{{\cal D}_{1}P_{k}}{\partial_{2} c}-
\frac{1}{2k}\frac{P_{k} \cdot {\cal D}_{1}(\partial_{2}
c)}{(\partial_{2}
c)^2}].
\label{r}
\end{equation}
By substituting
$a$ and $c$ from (\ref{10}) to these formulae we'll get
the following recursive differential equation
\begin{equation}
\partial_{2} b_{2k} -(
\frac{4ky_{1}(y_{1}^{2}-3y_{2}^{2})}
{\zeta(y_{1}^{2}+y_{2}^{2})}+\frac{1}{y_{2}})b_{2k}=L_{2k},
\label{Tb}
\end{equation}
where
\begin{equation}
L_{2k}=
-\frac{2y_{1}y_{2}}{\zeta} \cdot
[R_{k} +
\frac{y_{1}^{2}+y_{2}^{2}}{\zeta y_{2}}P_{k}-
\frac{(y_{1}^{2}-y_{2}^{2})}{2y_{1}y_{2}}
\cdot \frac{\partial_{2} P_{k}}{2k}-
\frac{\partial_{1} P_{k}}{2k}+
\frac{y_{1}^{2}+y_{2}^{2}}{y_{1}y_{2}^{2}}\cdot\frac{P_{k}}{4k}] ,
\label{Lk1}
\end{equation}

\begin{equation}
P_{k}=
\frac{4\zeta}{y_{1}^{2}+y_{2}^{2}}\cdot
(y_{1}{\partial}_{1}b_{2k-2}-
y_{2}{\partial}_{1}a_{2k-2})+
\frac{8(k-1)\zeta}{(y_{1}^{2}+y_{2}^{2})^{2}} \cdot
[(y_{1}^{2}-y_{2}^{2}) \cdot b_{2k-2}-
2y_{1}y_{2} \cdot a_{2k-2}],
\label{Pk1}
\end{equation}
$k\neq 1$ and
$
P_{1}=
16y_{1}y_{2}\zeta/(y_{1}^{2}+y_{2}^{2})^{2}$
and $R_{2k}$ is given by (\ref{Rk}) and
$
R_{1}=4y_{1}\zeta^{2}/(y_{1}^{2}+y_{2}^{2})^{2}.
$

>From (\ref{le1}) we get
\begin{equation}
a_{2k}=\frac{\partial _{1} c}{\partial _{2}c} b_{2k}-
\frac{1}{\partial _{2}c}\frac{P_{k}}{2k}.
\label{ar}
\end{equation}

\bigskip

{\it  Explicit formulae for $b_{2}$ and $a_{2}.$}
\bigskip

Let us demonstrate how this works,
in the first (nontrivial) order, when  the equations (\ref{le1}),
(\ref{le2}) take the form:

\begin{equation}
2{\partial}_{2} b \cdot a_{2} -
2{\partial}_{2} a \cdot b_{2} +\{c,a_{2}\}_{(1,2)}+
\{d,b_{2}\}_{(2,1)}=
4 c{\partial}_{1} c
-4 d{\partial}_{2} c,
\label{e12o}
\end{equation}

\begin{equation}
-{\partial}_{2} c \cdot a_{2}+
{\partial}_{1} c \cdot b_{2}=
- 2 c{\partial}_{1} b -
2 d{\partial}_{1} a.
\label{e22o}
\end{equation}

In accordance with (\ref{Tb}), (\ref{Lk1}) and (\ref{Pk1})
we have
\begin{equation}
\partial_{2} b_{2} -
(\frac{4y_{1}(y_{1}^{2}-3y_{2}^{2})}
{\zeta(y_{1}^{2}+y_{2}^{2})}+\frac{1}{y_{2}})b_{2}=
-\zeta \cdot \frac{8y_{1}^{2}y_{2}}{(y_{1}^{2}+y_{2}^{2})^{2}}
-\frac{1}{\zeta}\cdot\frac{32y^{2}_{1}y_{2}}{y_{1}^{2}+y_{2}^{2}}.
\label{bt}
\end{equation}
This is an ordinary differential equation that can be solved in
quadratures.

We get the following
\bigskip

{\bf Theorem.}~~
Let $\cal S$ be a three-surface:
\begin{equation}
w(x^{3}, \upsilon)= \frac{\zeta}{\upsilon}\tanh x^{3},~~~~~
s(x^{3},\upsilon,\bar{\upsilon})=-2 \ln{\upsilon}
+ \sum_{k=1}^{\infty}2^{2k} (\cosh{x^{3}})^{-2k}g_{2k}(\upsilon,
\bar{\upsilon}),
\label{1aa}
\end{equation}
where
$g_{2k}(\upsilon, \bar{\upsilon})= a_{2k}(y_{1},y_{2})+ib_{2k}
(y_{1},y_{2}),$
$\upsilon=y_{1}+iy_{2},$
$b_{2k}$ is a solution of recursive differential equations
(\ref{Tb}) and $a_{2n}$ is given by (\ref{ar}) then
$ \cal S$ satisfies $\Phi |_{{\cal S}}=dx^{3} \wedge dx^{4} \wedge
dx^{5}$ with the following boundary
conditions:
near
$x^{3}=- \infty $,
$\cal S $  looks like ${\bf R} \times \Sigma$,
with $\Sigma $ being the Riemann surface defined by
$w = \zeta {\upsilon }^{-1}$, $s=-2\ln \upsilon$;
near  $x^{3}=+\infty,~\cal S $  looks like
${\bf R} \times {\Sigma}^{'}$ with ${\Sigma}^{'}$
being  the Riemann surface defined by
$w=-\zeta \upsilon^{-1},  s=-2\ln \upsilon.$

In this theorem we don't discuss the convergence
of the series, i.e. we have constructed the surface as a formal series.
The problem of convergence will be discussed below.

\bigskip

{\it Remark} 1. {\it Large} $\zeta $

\bigskip

Equation (\ref{bt})
crucially simplifies if one sends $\zeta \to \infty$.
In fact sending $\zeta
\to \infty$, one gets the following equation for $b_{2}$:

\begin{equation}
\partial_{2} b_{2} -
\frac{b_{2}}{y_{2}}=
-\zeta \cdot \frac{8y_{1}^{2}y_{2}}{(y_{1}^{2}+y_{2}^{2})^{2}}.
\label{bzi}
\end{equation}

>From this ordinary differential equation one can easily
get the solution for $b_{2}$:
\begin{equation}
b_{2}=
-\zeta \cdot
\frac{4y_{2}(y_{1}y_{2}+(y_{1}^{2}+y_{2}^{2})\arctan{y_{2}/y_{1}}
)}
{y_{1}(y_{1}^{2}+y_{2}^{2})}
+\zeta \cdot f(y_{1}) y_{2},
\label{solb2}
\end{equation}
here $f(y_{1})$ is an arbitrary  function. Therefore in this regime
$b_{2}\sim \zeta$. According  (\ref{ar})  $a_{2}\sim \zeta$ and
\begin{equation}
a_{2}=
\frac{y_{1}^{2}-y_{2}^{2}}{2y_{1}y_{2}}b_{2}.
\label{aaa}
\end{equation}

Equation (\ref{Tb}) for arbitrary $k>2$ is also simplified and has
the form
\begin{equation}
\partial_{2} b_{2k} -\frac{1}{y_{2}}b_{2k}=L_{2k},
\label{Tbzi}
\end{equation}
Let us show that $b_{2k}$ for large $\zeta$ is proportional to
$\zeta$. For $k=1$ we have just proved it.
Under assumption that $b_{2r}\sim \zeta$
for $1<r<k$, we get that $P_{k}\sim \zeta ^{2}$,
$R_{k}\sim \zeta ^{2}$, $k>1$, so we neglect the second term in
 the square brackets in (\ref{Lk1})
and we left with
\begin{equation}
L_{2k}=
-\frac{2y_{1}y_{2}}{\zeta} \cdot
[R_{k} -\frac{(y_{1}^{2}-y_{2}^{2})}{2y_{1}y_{2}}
\cdot \frac{\partial_{2} P_{k}}{4k}-
\frac{\partial_{1} P_{k}}{2k}+
\frac{y_{1}^{2}-y_{2}^{2}}{y_{1}y_{2}^{2}}
\cdot\frac{P_{k}}{k}] ,
\label{Lk1aa}
\end{equation}
i.e. $L_{2k}\sim \zeta$ and we get $b_{2k}\sim \zeta $
and $a_{2k}\sim \zeta $, $k>0$.

\bigskip

{\it Remark} 2. {\it Small } $\zeta $
\bigskip

Let us consider the case when $\zeta \to 0$.
First let us examine the behavior of
$b_{2}$ for small $\zeta$.
For small $\zeta$ we leave $1/\zeta$ terms in the (\ref{bt})
and we get an algebraic relation
\begin{equation}
b_2=
\frac{8y_{1}y_{2}}{y_{1}^{2}-3y_{2}^{2}},
\label{zo1}
\end{equation}
here we assume that $y_{1}^{2}\neq 3y_{2}^{2}$.
Therefore we have
\begin{equation}
s(y_{3},\upsilon,\bar {\upsilon})= -2\ln \upsilon +
\frac{4}{(e^{y_{3}}+e^{-y_{3}})^{2}}\cdot
\frac{\upsilon{\bar \upsilon} -{\bar \upsilon^{2}}}{\upsilon^{2}+
{\bar \upsilon}^{2}
-\upsilon{\bar \upsilon}}+....
\label{bzoc}
\end{equation}

For arbitrary $k$ in the left hand side of (\ref{Tb}) one also leaves
only
$1/\zeta$ term
\begin{equation}
b_{2k}=\frac{y_{2}(y_{1}^{2}+y_{2}^{2})}{2k(y_{1}^{2}-3y_{2}^{2})}
[R_{k} +
\frac{y_{1}^{2}+y_{2}^{2}}{\zeta y_{2}}P_{k}].
\label{Tbzo}
\end{equation}
Since $P_{k}$ is proportional to $\zeta$
we have to keep the second  term in the right hand side of (\ref{Tbzo})
and we have  that $b_{2k}$ and $a_{2k}$ do not depend on $\zeta $
for small $\zeta$.

\bigskip

{\it  Remark} 3. {\it On the convergence of the series}
\bigskip

Let us make few comments about the convergence.
For this purpose let us examine the behavior of the system
(\ref{le1}) and  (\ref{le2}) for the large $k$.
For the large $k$ we can drop out the last two terms in the left hand
side of
(\ref{le2})
and get an algebraic system
\begin{equation}
-\partial _{2}c  \cdot a_{2k}+
 \partial _{1}c  \cdot b_{2k}=\frac{1}{2k}\cdot P_{k};
\label{le1a}
\end{equation}
\begin{equation}
-\partial_{1} a \cdot a_{2k}
+\partial_{2} a \cdot b_{2k}=\frac{1}{2k}\cdot R_{k}.
\label{le2a}
\end{equation}
For $k>r$ with  $r$ being large enough we have
from (\ref{ncl})

\begin{equation}
kW_{1} b_{2k} =L^{as}_{2k},
\label{ncla}
\end{equation}
where
\begin{equation}
L^{as}_{2k}=
 \frac {\partial_{2} c}{(\partial_{1} c)^2+(\partial_{2}c)^2}
 \cdot
[-R_{k} + P_{k}\frac{\partial_{1} a}{\partial_{2} c}].
\label{ra}
\end{equation}
More explicitly this can be written as
\begin{equation}
-\frac{4ky_{1}(y_{1}^{2}-3y_{2}^{2})}
{\zeta(y_{1}^{2}+y_{2}^{2})}b_{2k}=L^{as}_{2k},
\label{Tba}
\end{equation}
where
\begin{equation}
L^{as}_{2k}=
-\frac{2y_{1}y_{2}}{\zeta} \cdot
[R_{k} +
\frac{y_{1}^{2}+y_{2}^{2}}{\zeta y_{2}}P_{k}].
\label{Lk1a}
\end{equation}

Note that for the large $k$ in the left hand side of (\ref{Pk})
we can leave only the terms proportional to $k$ and
this gives
\begin{equation}
P^{as}_{k}=
8k({\partial}_{1} c \cdot b_{2k-2}-
{\partial}_{2} c \cdot a_{2k-2}),
\label{Pka}
\end{equation}
or from (\ref{Pk1})
\begin{equation}
P^{as}_{k}=
\frac{8k\zeta}{(y_{1}^{2}+y_{2}^{2})^{2}} \cdot
[(y_{1}^{2}-y_{2}^{2}) \cdot b_{2k-2}-
2y_{1}y_{2} \cdot a_{2k-2}],
\label{Pk1a}
\end{equation}
Taking into account (\ref{le1}) we conclude that
\begin{equation}
-\partial _{2}c  \cdot a_{2k}+
 \partial _{1}c  \cdot b_{2k}=
4( -\partial _{2}c  \cdot a_{2k-2}+
 \partial _{1}c  \cdot b_{2k-2}),
\label{pa}
\end{equation}
i.e.
\begin{equation}
-\partial _{2}c  \cdot a_{2k}+
 \partial _{1}c  \cdot b_{2k}=
4^k{\cal P},
\label{paa}
\end{equation}
where
\begin{equation}
{\cal P}_{r}=\frac{1}{4^{r}}(-\partial _{2}c  \cdot a_{2r}+
 \partial _{1}c  \cdot b_{2r}),
\label{paaa}
\end{equation}
Therefore we have
\begin{equation}
P^{as}_{k}=2k\cdot 4^{k}{\cal P}.
\label{ppaaa}
\end{equation}
Using (\ref{r}),
we have
\begin{equation}
a_{2k}=\frac{\partial _{1} c}{\partial _{2}c} b_{2k}-
\frac{1}{\partial _{2}c}\cdot 4^{k}{\cal P},
\label{ra1}
\end{equation}
Using (\ref{ar}) one can  represent the sum over $s$ in (\ref{Rk})
as
\begin{equation}
R_{k}= -2\partial _{2}(\partial _{1}c/\partial _{2}c)
\sum_{s=1}^{k-1}sb_{2s}b_{2k-2s}
-\sum_{s=1}^{k-1}P_{2s}\partial _{2}b_{2k-2s}
+\sum_{s=1}^{k-1}\frac{s}{k-s}b_{2s}\partial _{2}P_{2k-2s}.
 \label{rkes}
 \end{equation}
In the second and the third sum we can use the asymptotic relation
(\ref{ppaaa}) for $s>r$ and $k-s>r$, respectively.
So, schematically we can represent $R_{k}$ for the large
$k$ as
\begin{equation}
R_{k}\sim  \sum_{s1}^{k-1}[s\gamma b_{2s}b_{2k-2s}+
\lambda _{s} \partial_{2}b_{2k-2s}+
\omega _{s}b_{2k-2s} ],
\label{Rkex}
\end{equation}
$\lambda _{s}$ and  $\omega _{s}$ depend only on the first $r$ terms
$b_{2k}$
and $a_{2k}$; $\gamma =-2\partial _{2}(\partial _{1}c/\partial _{2}c)$.

So we have to solve the following problem:
to prove the convergence
of series  $\sum _{k>r} b_{2k} \tau^{k}$ in some circle, $|\tau|<\rho$
if we have the following recursive relations

\begin{equation}
kW_{1} b_{2k} =
 \frac {-2\partial_{2} c}{(\partial_{1} c)^2+(\partial_{2}c)^2}
 \cdot
[\sum_{s=r}^{k-1}s(\lambda _{s}
\partial_{2}b_{2k-2s}+
\omega _{s}b_{2k-2s})
+\gamma sb_{2s}b_{2k-2s}
 + \theta_{k}].
\label{abk}
\end{equation}
with uniformly bounded functions $\lambda _{s}(y_{1},y_{2})$,
$\omega(y_{1},y_{2})$  and $\theta_{s}(y_{1},y_{2})$,
$$|\lambda
_{s}|<\lambda ^{s},~~|\omega _{s}|<\omega ^{s},~~
|\theta_{s}|<\theta^{s}.$$
Since we deal with series (\ref{11a}) and (\ref{11b}),
in our case $\tau = \gamma (y^{3})$ and $|\gamma| \leq 1/2.$

To solve this problem it seems suitable to
introduce a generating function
\begin{equation}
\Psi ((y_{1},y_{2},\tau))=
\sum _{k>1} b_{2k}(y_{1},y_{2}) \tau^{k}
\label{gf}
\end{equation}
This generating function satisfies
the following partial differential equation
$$
\partial _{\tau}\Psi (y_{1},y_{2},\tau)+
\Gamma (y_{1},y_{2},\tau)\Psi (y_{1},y_{2},\tau)
\partial _{\tau}\Psi (y_{1},y_{2},\tau)
+\Lambda (y_{1},y_{2},\tau)\partial _{y_{2}}\Psi (y_{1},y_{2},\tau)+
$$
\begin{equation}
\Omega (y_{1},y_{2},\tau)\partial _{y_{2}}\Psi (y_{1},y_{2},\tau)=
\Theta (y_{1},y_{2},\tau).
 \label{degf}
 \end{equation}

The explicit formulae for $\Gamma (y_{1},y_{2},\tau)$,
$\Omega _{y_{2}}(y_1,y_2,\tau )$
 $\Lambda(y_1,y_2,\tau)$ and $\Theta(y_1,y_2,\tau)$
follow from equation (\ref{le1}), (\ref{ra1}) and (\ref{abk}).
We see that the  generating function satisfies the  nonlinear
partial differential equation  over
$y_2$ and $\tau$ of the first order.
The variable $y_{1}$ enters here as a parameter.
It is well known that under suitable conditions on
$\Omega$, $\Lambda$, $\Gamma$ and $\Phi$ the Cauchy problem for
equation (\ref{degf}) has an unique solution and the solution can
be obtained by finding the first integrals of equations for
characteristics.
\section {Conclusion}
In this note the equation for the domain wall in MQCD was derived and
some
properties of solution of this equation have been discussed.
The problem of the rigorous proof of the existence of the domain wall
 solution of the equation requires a further consideration.

Recently a different approach was suggested in \cite{Fayya}
where an explicit intersecting fivebrane
configuration was found and it was interpreted as  a domain wall in
MQCD.  Another approach to the problem of domain walls in MQCD has been
explored in \cite{STY}

 \section{Acknowledgments}

I am very grateful to Professor E. Witten for
important advises, correspondence and for pointing
out that parameter $\zeta$ was omitted
in the preliminary draft of the paper.
I also want to thank Professors A. A. Belavin
and D. V. Gal'tsov for attention to the work.
This work is partially supported by
Soros Student Fellowship, grant S97-602316.

\end{document}